# Massive Black Holes Across Cosmic Time

A White Paper for the Astro2010 Decadal Review

Lead Author: Piero Madau

Department of Astronomy and Astrophysics

University of California Santa Cruz

(831) 459-3839   pmadau@ucolick.org

Contributors: T. Abel (Stanford), P. L. Bender (U of Colorado), T. Di Matteo (CMU), Z. Haiman (Columbia), S. A. Hughes (MIT), A. Loeb (Harvard), E. S. Phinney (Caltech), J. R. Primack (UCSC), T. A. Prince (Caltech), M. J. Rees (IoA, Cambridge), D. O. Richstone (U of Michigan), B. F. Schutz (AEI, Potsdam), K. S. Thorne (Caltech), M. Volonteri (U of Michigan)

# Massive Black Holes Across Cosmic Time

The formation, assembly history, and environmental impact of the massive black holes (MBHs) that are ubiquitous in the nuclei of luminous galaxy today remain some of the main unsolved problems in cosmic structure formation studies. The recent observational breakthroughs achieved in this field by ground and space-based facilities – from the discovery of quasars at redshifts above 6 [1] to the tight correlation measured between hole masses and the stellar velocity dispersion of the host stellar bulge [2], from the presence of MBH binaries revealed in the nucleus of NGC 6240 [3] and the radio core of 3C 66B [4] to the discovery of the bimodality of galaxy colors [5] – have not yet been accompanied by equally significant progress in our understanding of the evolutionary history of these objects. Numerical simulations are not yet able to predict the formation path of the first MBH seeds at cosmic dawn. Close MBH binaries are expected to form in large numbers during cosmic evolution [6], but their merging history is *terra incognita*. The relative roles of gas and stellar dynamical processes in driving wide binaries to coalescence [7], and the effect of black hole mergers and gravitational wave recoil on nuclear stellar cusps and off-nuclear AGN activity [8,9,10], remain poorly understood. Accreting black holes can release large amounts of radiative and kinetic energy to their surroundings, and play a crucial role in determining the thermodynamics of the interstellar, intracluster, and intergalactic medium [11]. The detailed astrophysics of these processes and their importance in regulating e.g. star formation in the host galaxies remain open questions. The *WMAP* 5-year data

**Some key science questions in galaxy evolution studies addressed in unique ways by LISA**

| Key Science Question | Relevant LISA Capabilities |
|---|---|
| When and how did the universe emerge from its dark ages? | LISA can detect the coalescence of $10^4$ $M_\odot$ MBH binaries up to redshift 15. |
| Where are the remnants of the first stars today? | LISA can detect "Population III remnant" holes of 300 $M_\odot$ spiraling into more massive companions at redshifts up to 10. |
| How often do galaxy merge and how do mergers shape the Hubble sequence? | LISA can trace the merger history of $<10^7$ $M_\odot$ MBHs and their host galaxies during the "quasar era". |
| How do MBHs grow, "know" about their host stellar spheroids, and influence galaxy formation? | LISA can determine the mass and spin distributions in merging MBH binaries as a function of cosmic time. |

constrain the redshift of cosmological (instantaneous) reionization to be 11±1.4 [12], an indication that significant star-formation activity started at very early times. We infer that massive stars and accreting black holes must have been shining when the universe was less than 400 Myr old. But the impact they had on their environment and on the formation and structure of more massive systems remains an open question.

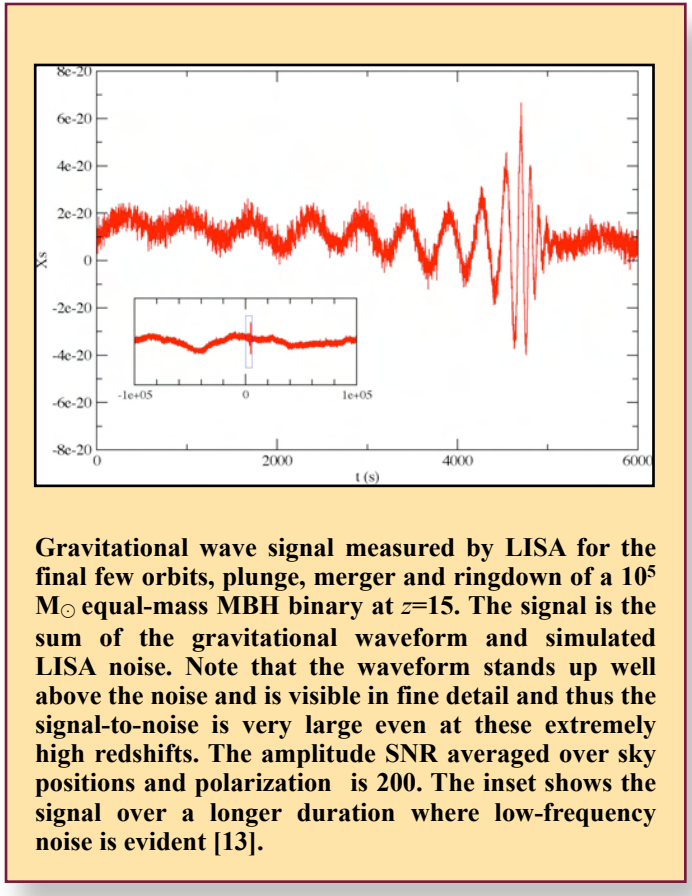

**Gravitational wave signal measured by LISA for the final few orbits, plunge, merger and ringdown of a $10^5$ $M_\odot$ equal-mass MBH binary at $z$=15. The signal is the sum of the gravitational waveform and simulated LISA noise. Note that the waveform stands up well above the noise and is visible in fine detail and thus the signal-to-noise is very large even at these extremely high redshifts. The amplitude SNR averaged over sky positions and polarization is 200. The inset shows the signal over a longer duration where low-frequency noise is evident [13].**

LISA will provide critical new information on the distances and rates of MBH binary coalescences including precision measurements of the mass and spin of the binary constituents across cosmic time. At LISA frequencies the strongest sources are mergers of MBH binaries in the mass range between $10^4$ and $10^7\,M_\odot$. The binaries start with wide orbits at low frequencies. As they lose energy their frequency increases and their radiation strengthens. A typical source enters the LISA band a year or more before the final merger so many orbits are recorded, encoding details of the system properties and behavior, position on the sky, and absolute distance. The coherent phase and polarization information obtained over LISA solar-orbit baseline (and variable orientation) can pinpoint a source in the sky to a few arcminutes in the best cases, and typically to better than a degree. In the last hours or minutes the signal-to-noise ratio grows very high, often into the hundreds to thousands depending on distance. At its peak luminosity a MBH binary is the most extreme converter of mass-energy in the universe, radiating a power $10^{50}$-$10^{51}$ watts in a few wave cycles. This exceeds by orders of magnitudes the luminosity of all the stars in the visible universe. MBH binary inspiral and coalescing events are such powerful radiators that LISA can detect them anywhere, out to the epoch of cosmological reionization, thereby providing an entirely new window on the distant universe.

## The coevolution of galaxies and MBHs from *z*=15 to *z*=0

LISA probes directly and intimately into the detailed assembly history of galaxies: the anticipated large samples of MBH binary mergers will provide a direct record of the whole history of structure formation in the observable universe, and of the astrophysical processes that grew MBHs in the nuclei of galaxies. LISA will explore a mass range that is complementary to that probed by the distant massive quasars ($>10^7$ $M_\odot$), providing a complete census of MBHs across cosmic time. The coevolution of MBHs and their host galaxies in hierarchical structure formation scenarios gives origin to a number of key questions that are at the core of LISA unique science capabilities:

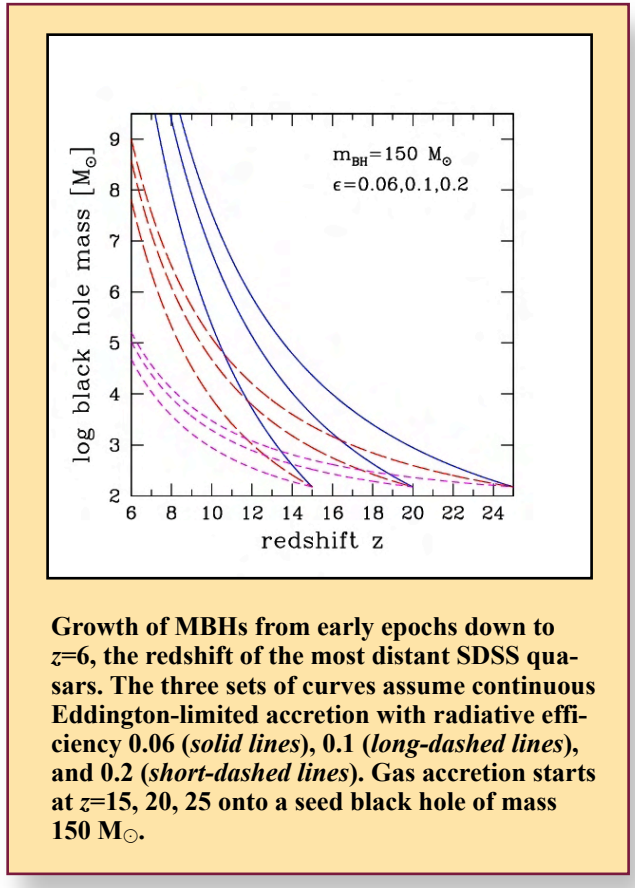

Growth of MBHs from early epochs down to *z*=6, the redshift of the most distant SDSS quasars. The three sets of curves assume continuous Eddington-limited accretion with radiative efficiency 0.06 (*solid lines*), 0.1 (*long-dashed lines*), and 0.2 (*short-dashed lines*). Gas accretion starts at *z*=15, 20, 25 onto a seed black hole of mass 150 $M_\odot$.

**Did the first MBHs form in subgalactic units far up in the galaxy merger hierarchy, well before the bulk of the stars observed today?** The seeds of the redshift 6 quasars discovered in the SDSS had to appear at very high redshift, *z*>10, if they start small (< $10^4$ $M_\odot$), and are accreting no faster than the Eddington rate (see Figure). In hierarchical clustering cosmologies the physical conditions (density, gas content) in the inner regions of mainly gaseous high-redshift protogalaxies appear to make them natural candidates as nurseries for seed black holes. The first MBHs may naturally arise from the collapse of the very first Population III stars. In this scenario, ~100 $M_\odot$ nuclear black holes began to form sometime before redshift 15, when gas in subgalactic halos cooled rapidly and fragmented [14]. An alternative, direct route to the formation of $10^5$ $M_\odot$ MBHs in the first galaxies involves the rapid collapse of metal-free gas irradiated by a strong UV flux (which suppresses $H_2$ formation and fragmentation into stars [15]). Early holes get incorporated through a series of galaxy mergers into larger and larger halos, sink to the center owing to dynamical friction, grow via gas accretion, and form binary systems. Seed holes accreting gas from the surrounding medium would shine as "miniquasars" at redshifts as high as 20, with dramatic effects on the thermodynamics of the intergalactic medium [16]. There are significant uncertainties about this key period in structure formation, from the fate of first stars [17,18] to the

growth of billion $M_\odot$ holes in redshift 6 quasars to the role of early black holes in the reheating and reionization of the universe [19,20,21]. It is certain, however, that early MBH seed formation is needed to account for the presence of bright quasars at early epochs. And while only a small fraction of the first seeds needs to grow supermassive by $z=6$, the larger substrate of sub-million $M_\odot$ holes are expected to evolve along the galaxy merger hierarchy and provide the bulk of the merger rate. The formation and evolution of the first MBHs is best traced by observing the coalescence of 300 to $10^5$ $M_\odot$ binaries detectable by LISA out to $z=10$, at a rate that will be very different in the two formation scenarios (slow growth of Pop III holes versus rapid direct collapse), and may be as high as 30 yr$^{-1}$ at $z=6$ [22].

**Is there a population of relic intermediate-mass black holes (IMBHs) lurking in low-redshift galaxies?** A clue to this question may lie in the numerous ultraluminous off-nuclear ("non-AGN") X-ray sources that have been detected in nearby galaxies. Assuming isotropic emission, the inferred masses of these "ULXs" may suggest IMBHs with masses above a few hundred $M_\odot$ [23]. Regardless of whether they are the seeds of the supermassive variety, LISA can search for the compact remnants of the first (Population III) stars, through observations of IMBHs spiraling into more massive companions at redshifts < 6. Detection of such "intermediate mass-ratio inspirals" (IMRIs) will provide information on the mass function of Population III stars, the growth, dynamical evolution, and fate of their remnants, and other mechanisms for forming IMBHs, such as the core collapse of globular clusters.

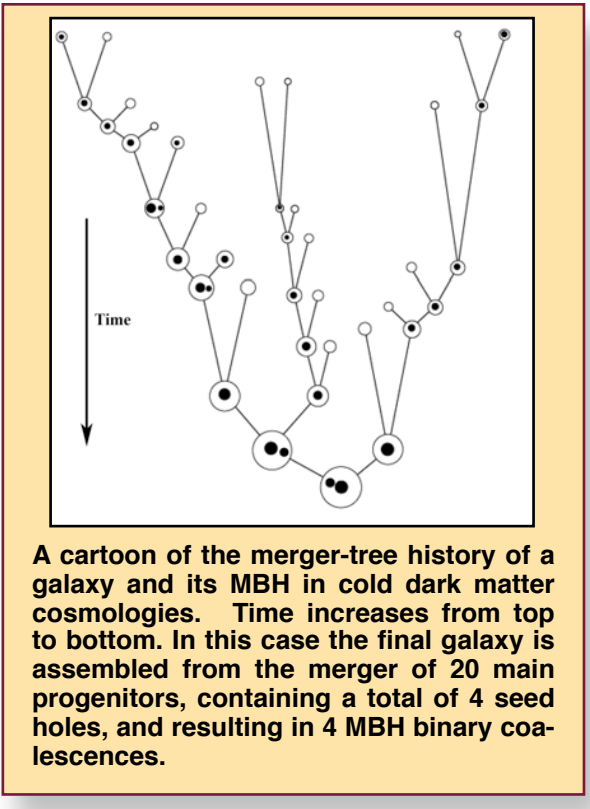

A cartoon of the merger-tree history of a galaxy and its MBH in cold dark matter cosmologies. Time increases from top to bottom. In this case the final galaxy is assembled from the merger of 20 main progenitors, containing a total of 4 seed holes, and resulting in 4 MBH binary coalescences.

**What is the merger history of MBHs and their host galaxies?** There is a simple argument bounding the number of mergers of MBHs that LISA is likely to see. The Hubble Space Telescope observes more than $10^{10}$ galaxies. Most bright local galaxies contain central supermassive black holes. Fossil evidence for mergers among local galaxies implies that about 70% of these have undergone a merger during the 8 Gyr since redshift 1 [24]. Therefore, the galaxy merger rate at $z < 1$ must be close to one per year. If the MBHs in merging galaxies merge in turn, the coalescence rate of MBHs should also be at least one per year. In cold dark matter cosmologies, however, present-day galaxies like the Milky Way grow by the merger of thousands of early-forming subunits. If each of these subunits initially contained a seed hole of $10^4$ M$_\odot$, the merger rate seen by LISA could be as high as one thousand per year! The actual rate is proportional to the fraction of protogalactic fragments that contain seed black holes massive enough for LISA to detect their mergers, multiplied by the fraction of galaxy mergers that lead to black hole coalescences [22,25]. Observationally, the paucity of active MBH pairs may point to binary lifetimes far shorter than the Hubble time, indicating rapid inspiral of the holes down to the domain where gravitational waves lead to their coalescence. Some models predict many tens of detectable mergers at $z<5$ in a 5-year mission, and several times more at earlier epochs. Since all bright galaxies at low redshifts appear to host MBHs, LISA will detect all galaxy mergers that ultimately lead to the formation of ~$10^6$ M$_\odot$ MBH binaries. The higher mass mergers are the strongest sources in the LISA band, making precise measurements of system parameters possible [26]. The determination of sky position to a few arc minutes together with the luminosity distance to less than a few percent (at $z<1$, limited by weak-lensing errors [27]) will facilitate the search for electromagnetic counterparts of the merger event among the ~$10^4$ galaxies in the error box. Electromagnetic counterparts will allow determination of the distance-redshift relation to higher accuracy than any other known method.

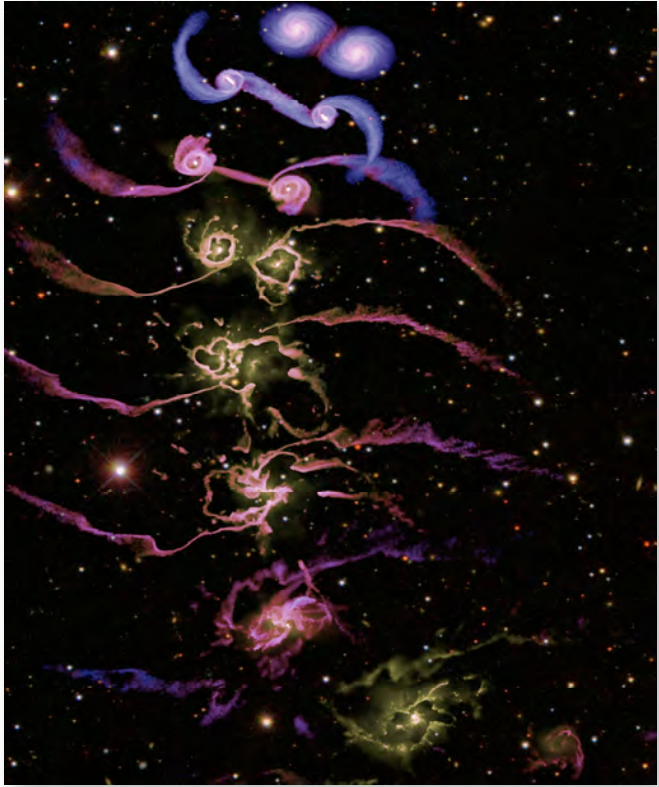

Snapshots from a simulation of a collision of two spiral galaxies, similar to the Milky Way, containing $10^5$M$_\odot$ MBHs at their centers. The images show only the gas in the galaxies. Color indicates temperature and brightness gas density. The collision drives both star formation in the galaxy and gas accretion onto the black holes, causing them to merge. The resulting quasar expels most of the gas from the galaxy, leaving a gas-poor galaxy containing a ~ $10^8$M$_\odot$ black hole (from [11]).

**What is the spin distribution of MBHs?** The spin of a MBH determines the efficiency of converting accreted mass into radiation and has implications for the direction of jets in active nuclei [28]. The electromagnetic braking of a rapidly spinning black hole may extract rotational energy [29], convert it into directed Poynting flux and electron-positron pairs, and power some radio galaxies and gamma-ray bursts. The orientation of the spin is thought to determine the innermost flow pattern of gas accreting onto Kerr holes [30]. The coalescence of two spinning black holes in a radio galaxy may cause a sudden reorientation of the jet direction, perhaps leading to the so-called "winged" or "X-type" radio source [31]. The spins of both quiescent and active MBHs are unknown. The distribution of spins will be a strong diagnostic of the mechanisms of black hole growth [32,33,34]. Recent breakthroughs in numerical relativity have provided us with a quantitative understanding of the spins resulting from comparable-mass binary hole mergers. LISA observations can provide information on the spins of the two MBHs in a binary prior to merger (inspiral) to within 1% [26] and on the spin of the merger remnant (ringdown). LISA will also measure orbital parameters (inclinations, eccentricities) and redshifted masses [(M(1+z)] extremely precisely (with less than percent level errors). Together with redshift measurements to ~10% accuracy, LISA will make it possible to build catalogs of mass/spin versus redshift to map out the evolution of MBHs with cosmic time.

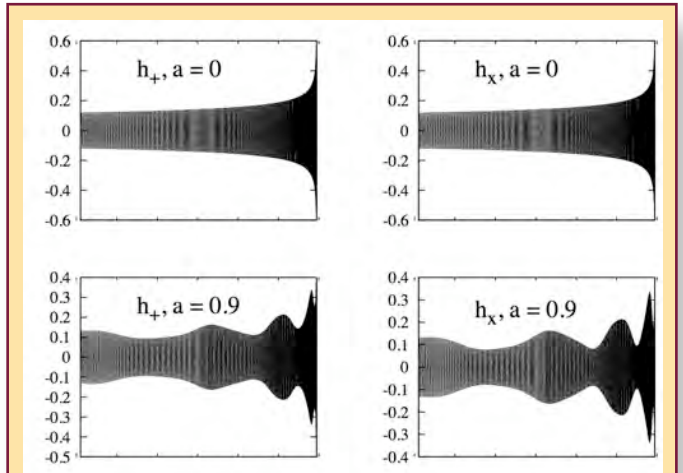

**The impact of black hole spin on the gravitational waves a binary produces. The top two panels show the waves generated in the coalescence of a pair of nonspinning black holes; the bottom two are the waves generated when each hole spins at 90% of the maximum allowed value. The strong modulations imposed by spin effects allow LISA to measure spin with exquisite accuracy [35].**

### References

[1] Fan, X. 2006, *NewAR*, **50**, 665

[2] Tremaine, S., et al. 2002, *ApJ*, **574**, 740

[3] Max, C. E., Canalizo, G., and de Vries, W. H. 2007, *Science,* **316**, 1877

[4] Rodriguez, C., et al. 2006, *ApJ*, **646**, 49